\documentclass[aps,prb,twocolumn,superscriptaddress,floatfix,eqsecnum,showpacs]{revtex4}
\usepackage{wrapfig}
\usepackage{subfigure}
\usepackage{psfrag}
\usepackage{tabularx}
\usepackage[dvips]{graphicx}
\DeclareGraphicsExtensions{.eps}

\begin{document}
\title{Metastable anisotropy orientation of nematic quantum Hall fluids}

\author{Daniel G. Barci}
\affiliation{Departamento de F{\'\i}sica Te\'orica
Universidade do Estado do Rio de Janeiro.
Rua S\~ao Francisco Xavier 524,
20550-013,  Rio de Janeiro, RJ, Brazil.}
\altaffiliation{Research Associate of the Abdus Salam International Centre for Theoretical Physics, ICTP, Trieste, Italy}

\author{Zochil Gonz\'alez Arenas}
\affiliation{Instituto de Cibern\'etica, Matem\'atica y F\'\i sica (ICIMAF)\\
Calle 15 \# 551  e/ C y D, Vedado, C. Habana, Cuba.}

\date{\today}

\begin{abstract}
We analyze the experimental observation of  metastable anisotropy
resistance orientation at half filled quantum Hall fluids by means
of a model of a quantum nematic liquid in an explicit symmetry
breaking potential. We interpret the observed ``rotation'' of the
anisotropy axis as a process of nucleation of nematic domains and
compute the nucleation rate within this model. By comparing with
experiment, we are able to predict the critical radius of nematic
bubbles, $R_c\sim 2.6 \mu m $. Each domain contains about $10^4$
electrons.
\end{abstract}



\maketitle

\section{Introduction \label{Introduction}}

{\em Quantum liquid crystals} are gapless  condensates  that
spontaneously break rotational and/or translational symmetry. There
is by now  a large amount of theoretical and experimental work
studying these new phases of strongly correlated systems in
different realizations like quantum hall systems\cite{F-K}, high
$T_c$ superconductors\cite{Nature} and heavy fermion
compounds\cite{Grigera}.

The two dimensional quantum smectic\cite{smectic}, also referred to
as stripe phase, is a metallic state that breaks translation
invariance in one direction.  It was conjectured that this modulated
electronic configuration is a good ground state of two dimensional
electron gases (2DEG)  under specific values of an external magnetic
field\cite{Fogler,chalker,F-K,Fisher-McDonald,QHSExp,QHSExp2}. In
fact, at partial filling factor the system tends to separate into
homogeneous fluids with different densities. Coulomb repulsion
frustrates this tendency and the system is forced to rearrange
itself by  lowering its dimensionality\cite{QHSI,LF}.
Collective excitations of this ground state were computed and anisotropic gapless fermionic
correlations, very different from those in the usual theory of Fermi
liquids\cite{QHSI, LF,QHS2, Lopatnikova,Fertig}, were
obtained.

Strong thermal, as well as quantum fluctuations of the stripes,
could produce topological defects, dislocations or disclinations
that, under appropriate circumstances, can melt the stripe order
into an homogeneous but anisotropic liquid\cite{Wexler}. This new
state is called the quantum nematic state\cite{Vadim,Lawler}, and it
is probably the best candidate\cite{Manusakis1,Manusakis2} to
explain the anisotropies observed in 2DEG at half filled Landau
levels\cite{QHSExp,QHSExp2}. In fact, experiments are compatible
with the interpretation of a spontaneous rotational symmetry
breaking at approximately $150 mK$ and a weak native potential,
responsible for aligning the principal axis resistance, of the order
of $1 mK$ per electron\cite{native}. On the other hand, no pinning
was detected in the $I-V$ curves, which suggests a liquid
state rather than periodical arrays.

Some models\cite{Yang} were proposed to understand the native
symmetry breaking potential responsible for the alignment of the
anisotropy. However, the origin of this potential remains
unknown. In a recent experiment\cite{Eisenstein}, the structure of
the native potential was studied on highly mobility samples over a
large scale of temperatures and magnetic fields. It was
reported a non trivial behavior of the resistance anisotropy as a
function of temperature and filling factor. It was also
found that the ``easy'' direction (the direction with lower
resistivity) can be aligned along the $\langle 1\bar{1}0\rangle$  or
the    $\langle 110\rangle$ crystallographic axes of the host $GaAs$
structure. The actual direction preferred by the system 
depends on the filling factor and on the in-plane magnetic field.
These directions can be interchanged according to the magnetic field
sweep in such a way that an interesting hysteresis pattern, typical
of metastability, comes up close to half filling. Moreover, to confirm the
picture of a bi-stable potential, Cooper {\em et.\ al.\
}\cite{Eisenstein} were able to quench the system in a  metastable
state (say the ``easy'' direction along  $\langle 110\rangle$).
Then, they observed the slow relaxation to the equilibrium state,
aligned with  the axis $\langle 1\bar{1}0\rangle$, for several final
temperatures.

In this work, we  analyze this result in the framework of a nematic
quantum fluid, submitted to a two component external potential; one
component with nematic symmetry and the other one with
tetragonal symmetry, possibly induced by the host $GaAs$
structure\cite{piezo}. For this purpose, we introduce an $XY$ model,
describing an effective quantum nematic phase, with a general
external symmetry breaking potential that produces a two orthogonal
minima structure. Within this picture, we assume that the decay of
the metastable state can be produced by  thermal activation over a
potential barrier. We expect that bubbles nucleation of the true
ground state into the metastable state, produced by long wavelength thermal
fluctuations, is the main mechanism responsible for the decay. This assumption is reasonable
provided the energy of the critical bubble is much greater than the
equilibrium temperature. We will show that this is, in fact, the
case for the data of ref. \onlinecite{Eisenstein}.

We use the Langer\cite{Langer} homogeneous nucleation theory  to
compute decay rates of the metastable state. We calculate the
critical bubble profile of the model, and estimate the critical
energy and radius by using two methods: an analytical
variational approach and an exact numerical computation.  By
comparing with the experiment, we are able to predict the radius of
critical domains of the order of $2.6 \mu m$, containing
approximately $10^4$ electrons. We find that the time evolution of
the anisotropy resistance observed in  ref. \onlinecite{Eisenstein}
is in agreement with the picture of nucleation of nematic domains,
with  directors pointing along a
stable direction, in a metastable nematic background with the
principal axis aligned in the perpendicular direction. We also show
that the homogeneous nucleation theory is good enough to make
estimations at first order, while the thin wall approximation is not
quite accurate. Our calculations predict that the domains have broad
walls in the actual experimental conditions.

We present our model of quantum nematic in an external potential in
section \ref{model}. To make the paper self contained and to fix
notations, we briefly review the theory of two-dimensional
homogeneous nucleation in our context in \S\ref{nucleation}. Then,
in \S \ref{Nematicnucleation}, we compute the decay rate, the
critical energy and the radius of the critical bubble as a function
of the parameters of our model, by using a variational
approach and the thin wall approximation. To check our
approximations, we numerically integrate the differential equation
which defines the critical bubble and compare the results with our
previous analytical estimations in section \ref{numerical}. Finally,
we compare our analysis with experimental results in
\S\ref{experiments} and summarize our conclusions  in section
\ref{summary}.

\section{$XY$ model of a quantum nematic liquid in a symmetry breaking potential \label{model}}

A nematic state is a homogeneous orientational ordered state with
the forward and backward directions identified.  That means that if
the system has a preference axis orientated along an angle $\theta$,
the state has the nematic symmetry $\theta \longrightarrow \theta
+\pi$. In two dimensions, this property is encoded in the definition
of the complex order parameter $Q=\rho\; e^{i2\theta}$, where the
argument $2\theta$ guarantees the nematic symmetry. Close
to the isotropic-nematic transition, we can expand the free energy
in powers of $Q$,
\begin{eqnarray}
F(Q)&=& \frac{1}{2}\;\int d^2x\;  \vec\nabla Q\cdot \vec\nabla Q^*+ \nonumber \\
&+&\frac{1}{V}\int d^2x\left\{ \frac{1}{2}a_2 Q Q^* + \frac{1}{4} a_4 (Q Q^*)^2 \right\} + \nonumber\\
&+& V(hQ)+\ldots\;,
\label{Fcomplex}
\end{eqnarray}
where $a_2$ and $a_4$ are arbitrary constants which depend
on the microscopic details of the system. $V(hQ)$ is an explicit
symmetry breaking term which depends on some external field $h$.
Differently from three dimensions, in the two dimensional
Ginzburg-Landau expansion of the nematic order parameter there is no
cubic term.

In the absence of external symmetry breaking potential, the modulus
of the order parameter spontaneously gets a non zero value $\rho$
 for  $a_2<0$,  while the angle $\theta$ remains
arbitrary. In two dimensions, the angle fluctuations are
logarithmically divergent\cite{KT}. Therefore, in the
absence of external fields there is no true order, but algebraically
decay quasi-long range order. However, in the presence of a small
symmetry breaking potential, this divergence is removed. In this
case, even small values of the external potential can produce big
values of the order parameter due to the huge susceptibility of this
transition.

The dynamics of the lowest energy modes is governed by the coarse-grained Hamiltonian 
(as usual, we have assumed at low temperatures a constant modulus of the order 
parameter in eq.~(\ref{Fcomplex}))
\begin{equation}
H=\int d^2x\;\;  \frac{J}{2}\; \left| \vec \nabla \theta \right|^2
+V(\theta)\;,
\label{Hamiltonian}
\end{equation}
where $J$ is the typical energy scale of the Kosterlitz-Thouless\cite{KT} phase transition, and $V(\theta)$ is an arbitrary 
potential that explicitly breaks rotation invariance, but preserves the nematic symmetry in 
such a way that $V(\theta)=V(\theta+\pi)$.

It is possible to parametrize this potential in terms of its even Fourier coefficients
\begin{equation}
V(\theta)=\sum_n\; h_{2n}\; \cos(2 n\theta)\;,
\label{potential}
\end{equation}
where $h_2$ is related with a nematic external field,  $h_4$ is a  symmetry breaking coefficient 
with tetragonal symmetry and so on.

This model, with $n=1$ in eq.~(\ref{potential}), was used to fit  the isotropic/anisotropic 
transition of a Hall liquid at $\nu=9/2$ filling factor, by means of  Monte Carlo simulations\cite{Manusakis1}. 
It was  shown that the general picture of a nematic liquid in a small symmetry breaking field  correctly describes the transition.
To use this model for fitting experimental data, it  is necessary to relate the nematic order parameter with observables. In ref.\  \onlinecite{Manusakis1}, it was shown that the relation
\begin{equation}
\frac{\rho_{xx}-\rho_{yy}}{\rho_{xx}+\rho_{yy}}= \langle \cos 2\theta\rangle + \ldots
\label{anisotropy}
\end{equation}
is a good approximation over a huge temperature range, except at extremely low temperatures where quantum fluctuations become important.
In eq.~(\ref{anisotropy}), $\rho_{xx}$ and $\rho_{yy}$ are the  measured longitudinal resistivities in the $x$ and $y$ direction respectively.

In this paper, we adopt the same criteria and study higher harmonics of the external field.
For this purpose, we  will analyze the effect of the second term $n=2$ in eq.~(\ref{potential}). Thus, we will consider a potential of the form
\begin{equation}
V(\theta)= h_2\; \cos(2\theta)-h_4\; \cos (4\theta)\;,
\label{potential2}
\end{equation}
where $h_2>0$ and $h_4>0$ are coefficients that measure the relative weights of the nematic and tetragonal components of the host symmetry breaking potential. It is interesting to note that, while the term $h_2$ is somehow mysterious, the term $h_4$ is not prohibited by symmetry and could be induced by the square structure of the host $GaAs$\cite{piezo}.

\begin{figure}[t]
\centering
\includegraphics[width=0.45\textwidth]{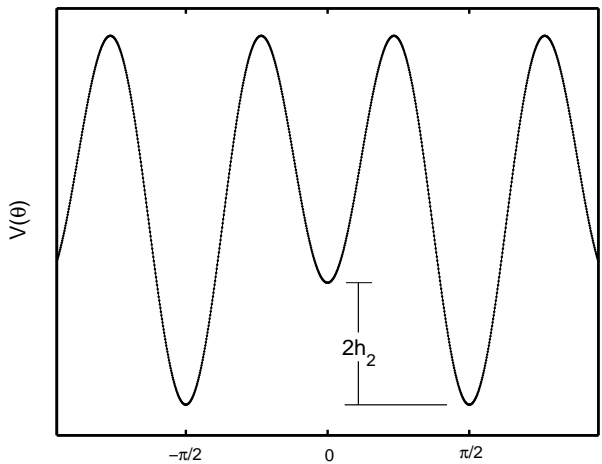}
 \caption{Potential of eq.~(\ref{potential}) with $h_2<4 h_4$. The metastable states are at $\theta= n \pi$, while the true ground state is align with $\theta=\pi (2n+1)/2$ with $n=0,\pm 1\ldots$.  Maxima are taken at $\cos(2\theta_{Max})=h_2/(4h_4)$. The energy difference between minima is $2h_2$ and  the height of the barrier
 in related with $h_4$ through eq.~(\ref{height}).}
  \label{pot}
\end{figure}

The net effect of the new term $h_4$ is to introduce local
metastable minima for the angular variable $\theta$. We depict the
potential for typical values of $h_2< 4 h_4$ in fig.~\ref{pot}. The
sign of $h_4$ controls which one of the different minima is
metastable and which one is the true ground state. We arbitrarily
choose this sign in such a way of having a metastable state
at $\theta= n \pi$  and the ground state at $\theta=\pi (2n+1)/2$
with $n=0,\pm 1\ldots$.   On the other hand, the potential
has maxima at $\cos(2\theta_{Max})=h_2/(4h_4)$.

The energy difference between the stable and metastable minima
is $2 h_2$, while $h_4$ is related with the height of the
energy barrier, in fact,
\begin{equation}
V(\theta_{Max})-V(0)= \frac{h_2^2+2h_4^2+h_2 h_4}{h_4}\; ;
\label{height}
\end{equation}
which in the limit of quasi degenerate minima reduces to  $V(\theta_{Max})-V(0)= 2 h_4$.

With this potential it would be possible, by a quenching process,  to prepare the system in a metastable phase; for instance, a nematic state with principal axis pointing in the $\theta=0$ direction. In these conditions, we would expect that long wavelength thermal fluctuations could produce nucleation of nematic domains with principal axis in the perpendicular direction $\theta=\pi/2$. Therefore, we should see that the anisotropy of the resistivity   ``rotates'' or changes in time from one direction to its perpendicular one as it has been observed\cite{Eisenstein}.

\section{Two dimensional  bubble nucleation \label{nucleation}}

One of the possible mechanisms for thermal activated decaying is the
nucleation\cite{Sahni} of  ground state domains in a homogeneous
metastable state which fills  all the available
area. The dynamics is determined by the domain's energy
that, in general, is a competition between a bulk contribution
(proportional to its area) and a boundary term (proportional to its
perimeter).

\begin{figure}
  \centering
\includegraphics[width=0.29\textwidth]{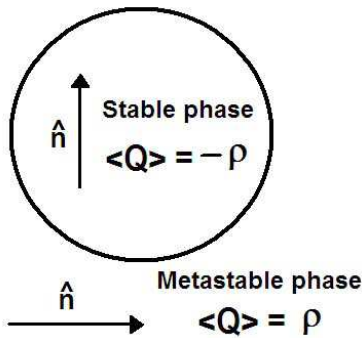}
 \caption{Sketch of a symmetrical bubble which  contains a nematic liquid with the director pointing in the $\theta=\pi/2$ direction. It is  embedded in a metastable state composed of a nematic liquid with the director pointing to $\theta=0$. $\rho$ is the modulus of the nematic order parameter and {\bf \^n} is the director.}
  \label{nucl}
\end{figure}

Consider,  for instance,  a bubble with spherical symmetry of radius
$R$ as depicted in fig.\ \ref{nucl}. In the thin wall approximation,
that is, when the width of the wall is much smaller than the radius,
the boundary and bulk contributions to the energy are well defined,
\begin{equation}
E(R)=-\pi \Delta{\cal F}\; R^2+ 2\pi \sigma \;  R +\ldots \; ,
\label{Energy}
\end{equation}
where $\Delta{\cal F}$  is the energy difference
between the stable and metastable states per unit area, $\sigma$ is
the surface tension, and the ellipsis  indicates subleading order in
the thin wall approximation.

While for small radius the positive boundary term dominates the
energy, for large $R$, the negative bulk contribution dominates. Due
to this competition, there is a critical radius
$R_c=\sigma/\Delta{\cal F}$, where the energy has a maximum
$E_c=\pi\sigma^2/\Delta{\cal F}$, and the critical  bubble is at
unstable equilibrium. The supercritical bubbles ($R>R_c$) will grow
until filling all the area with the ground state. On the other hand,
the subcritical bubbles ($R<R_c$) will shrink and finally disappear.
Both contributions are important in a  phase transition
since the actual mechanism is given by random long wavelength
thermal fluctuations which generate all types of bubbles. Some of
them will grow and others will shrink. In this picture,  the
transition is completed when the true ground state percolates  the
metastable one. The relative importance of these contributions
depends on the probability of fluctuations and on the growth rate of
the supercritical bubbles
\begin{equation}
\Omega=\frac{d~}{dt}\left[ \ln\left| \frac{R(t)}{R_c}-1\right|\right] \;.
\end{equation}

The important quantity to study nucleation is the  nucleation rate per unit of area. In the homogeneous nucleation theory of Langer\cite{Langer}, this quantity is given by $\Gamma=\Omega {\cal D}\; \exp(-E_c/T)$,
where $E_c$ is the energy of a critical bubble, $T$ is the final equilibrium temperature, $\Omega$ is the growth rate of a slightly supercritical bubble\cite{Alamoudi} and the prefactor ${\cal D}$ comes from the computation of fluctuations around the critical bubble profile.

The computation of $\Omega$ and ${\cal D}$ from microscopic quantum models is a very difficult task (see ref.\ \onlinecite{Alamoudi} and references therein).  However, it was shown\cite{Voloshin} that in  two dimensions and in the thin wall approximation ($\Delta{\cal F}\to 0$) these quantities can be cast in terms of the macroscopic parameter  $\Delta{\cal F}$,

\begin{equation}
\Gamma= \frac{\Delta{\cal F}}{2\pi\hbar}\;\; e^{-E_c/T} \;,
\label{Gamma}
\end{equation}
Interestingly, due to its two-dimensional character, this expression has no  corrections in powers of the thin wall adimensional parameter\cite{Voloshin} $\Delta{\cal F} T/\sigma^2$.

Homogeneous
nucleation theory is reliable provided  $E_c >> T$. For $E_c\sim T$, small amplitude thermal fluctuations could
trigger the phase transition and nucleation and spinodal decomposition can no longer be
distinguished.

Provided the conditions for homogeneous nucleation are satisfied, the probability of nucleating several bubbles at the same time is very small. We can estimate the typical time to complete a transition as the time of a simple nucleation event. Then,
\begin{equation}
\tau=\frac{1}{\Gamma A}= \frac{2\pi \hbar}{\Delta{\cal F} A}\;e^{E_c/T} \; ,
\label{tau}
\end{equation}
where $\tau$ is the time estimated to complete the transition, $\Gamma$ is the nucleation rate per unit of area, and $A$ is the total area of the sample considered.

This ``static'' approach will be sufficient for the purpose of this paper. However, time dependent corrections to Langer theory can be evaluated using out of equilibrium Shwinger-Keldish techniques\cite{Alamoudi}.

In the next section we will compute the time $\tau$ as well as the coefficients $\Delta{\cal F}$ and $\sigma$ in terms of the parameters of our model, $J$, $h_2$ and $h_4$.

\section{Nematic critical bubbles and the thin wall approximation \label{Nematicnucleation}}
A critical bubble is a radially symmetric static field configuration that solves the following differential equation
\begin{equation}
\nabla^2\theta-\frac{1}{J}\frac{\partial
V(\theta)}{\partial\theta}\,=\,0 \, ,\label{bubble}
\label{eqcriticalbubble}
\end{equation}
with the boundary conditions
$\lim_{r\to\infty}\theta(r)=0$ and
$\lim_{r\to\infty}\theta'(r)=0$. $V(\theta)$ is the potential of eq.~(\ref{potential2}) (fig.~\ref{pot}).

\begin{figure}
  \centering
\includegraphics[width=0.43\textwidth]{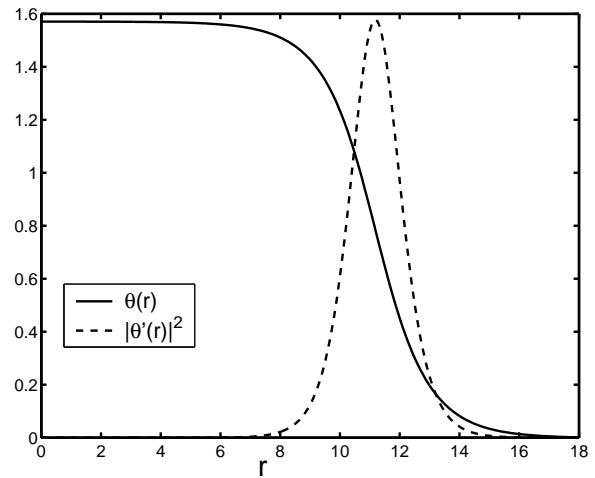}
 \caption{Typical radial symmetric bubble profile that solves eq.~\ref{eqcriticalbubble}. The continuous line represents the bubble configuration which starts close to the true ground state $\theta=\pi/2$ and reaches  the metastable state $\theta=0$ at asymptotically large radial distance. The dash line is the square derivative (properly rescaled) that defines the critical radius and the wall thickness}
  \label{solution}
\end{figure}

Such a solution describes a ``bubble'' like configuration which
starts close to the true ground state $\theta=\pi/2$ and reaches the
metastable state $\theta=0$ at asymptotically large distances. The
change from the stable to the metastable state occurs around the
critical radius $R_c$, over a distance $\xi$ which defines
the wall thickness of the bubble. We depict a typical profile of a
critical bubble and we also draw the square derivative of the
profile in fig.~\ref{solution}. The maximum of the derivative
defines the critical radius and the width of the peak is a measure
of the bubble wall thickness that, as we will show, is  related to
the nematic correlation length in the metastable phase.

Eq.~(\ref{eqcriticalbubble}) is an extremely difficult differential
equation to solve analytically and we will show a numerical
treatment in the next section. However, it is possible to have some
insight of its behavior through a variational analysis. The
idea is to propose a reasonable ansatz for the solution, by
considering the critical radius and the wall thickness as
variational parameters. Then, we determine these parameters by
extremizing the critical energy. We make the following ansatz,
\begin{equation}
\theta_b(r)=\frac{\pi}{4}
\left(
1-\tanh\left[ \frac{r-R_c}{\xi} \right]
\right) \, ,
\label{ansatz}
\end{equation}
where the radius $R_c$  and the wall thickness $\xi$ will be determined extremizing the energy.

\begin{figure}
  \centering
\includegraphics[width=0.45\textwidth]{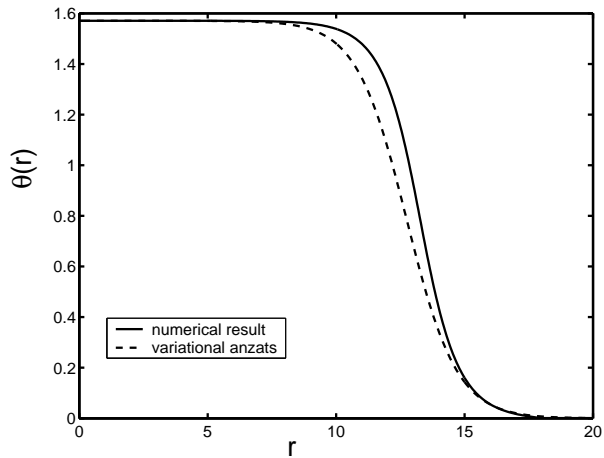}
 \caption{Comparison between a numerical solution of eq.~(\ref{eqcriticalbubble}) (continuous line) and the variational ansatz eq.~(\ref{ansatz}) (dash line).}
  \label{fig:ansatz}
\end{figure}

The form of this function is inspired in the problem of an
asymmetric quartic potential\cite{YuLu}. In that case,
eq.~(\ref{ansatz}) is the exact solution of
eq.~(\ref{eqcriticalbubble}) for the one-dimensional problem and,
for higher dimensions, it is the correct form in the thin wall
approximation $\xi/R_c<<1$\cite{Alamoudi}. Although our potential is
much more complicated than a simple quartic one, we expect to grasp
the general behavior and the correct order of magnitude with this
ansatz. Of course,  a variational technique is not a well controlled
approximation, thus,  in the next section, we compare our
variational analytical results  with a  numerical treatment of
eq.~(\ref{eqcriticalbubble}). Just to have some feeling on the
approximation,  we compare the variational profile of
eq.~(\ref{ansatz}) with the exact one, obtained from direct
numerical integration of eq.~(\ref{eqcriticalbubble}), in
fig.~\ref{fig:ansatz}.

Substituting  eq.~(\ref{ansatz}) into the Hamiltonian
eq.~(\ref{Hamiltonian}), we have the energy of the bubble
as a function of the variational parameters,
\begin{equation}
E(R_c,\xi)=\int d^2x\;\;  \frac{J}{2}\; \left| \vec \nabla \theta_b(r,R_c,\xi) \right|^2
+V(\theta_b((r,R_c,\xi))).
\label{BubbleEnergy}
\end{equation}
Integrating to leading order in $\xi/R_c$ we find an expression similar to eq.~(\ref{Energy}),
\begin{equation}
E(R_c,\xi)=-\pi \Delta{\cal F} R_c^2+ 2\pi \sigma(\xi)  R_c + O\left((\xi/R_c)^2\right)\; ,
\label{Energy1}
\end{equation}
where
\begin{equation}
\Delta{\cal F}= V(0)-V(\pi/2)= 2\; h_2 \\
\label{deltaF}
\end{equation}
essentially comes from the integral of the potential. On
the other hand, $\sigma(\xi)$ has contributions from both
terms of the integral and is given by
\begin{equation}
\sigma(\xi)=\frac{\pi^2}{32}\left(\frac{\pi}{2}\;\frac{J}{\xi}+0.39 V''(0)\; \xi \right).
\label{sigmaxi}
\end{equation}
The first term comes from the gradient contribution and  grows when
the thickness narrows. On the contrary, the second term,
which comes from the potential, is a linear increasing
function of $\xi$. Therefore, we fix this parameter looking for a
stationary solution
\begin{equation}
\frac{\partial E(R_c,\xi)}{\partial \xi} = 2\pi R_c \frac{d\sigma}{d\xi}=0 \; ,
\end{equation}
and obtain the optimal value,
\begin{equation}
\xi=2\; \sqrt{\frac{J}{V''(0)}} = \frac{J^{1/2}}{\sqrt{4h_4-h_2}} \; ,
\end{equation}
which  is two times the correlation length on the metastable phase.
With this value for the thickness, the superficial tension gets the simplest form,
\begin{equation}
\sigma\sim \frac{J}{\xi}= J^{1/2}\sqrt{4h_4-h_2} .
\end{equation}

Now,  we can  determine the critical radius $R_c$ by imposing
\begin{equation}
\frac{\partial E(R_c,\xi)}{\partial R_c} = -2\pi R_c \Delta {\cal F}+2\pi \sigma(\xi)=0.
\end{equation}
We have, in this way,
\begin{equation}
R_c=\frac{\sigma}{\Delta{\cal F}}= \frac{J^{1/2}}{2}\frac{\sqrt{4h_4-h_2}}{h_2}.
\label{eqRc}
\end{equation}
Finally, plugging   $\Delta{\cal F}$, $\sigma$ and $R_c$ into eq.~(\ref{Energy1}), we obtain the energy of  the critical bubble,
\begin{equation}
\frac{E_c}{J}=\frac{\pi}{2} \left\{ \frac{4h_4}{h_2}-1\right\}.
\label{Ecthinwall}
\end{equation}

It is necessary to have in mind that the thin wall approximation also impose restrictions  on the values of $h_2$ and $h_4$, since
\begin{equation}
\frac{\xi}{R_c}=\pi\frac{J}{E_c}=2\left(\frac{4h_4}{h_2}-1\right)^{-1}<<1.
\label{thinwall}
\end{equation}

This completes our estimation of relevant dynamical quantities in terms of the parameters of our model in the thin wall approximation.

\section{Numerical calculations \label{numerical}}

In order to check the range of validity of our variational approach
we numerically evaluate the bubble profile and the energy.

Assuming a solution with radial symmetry and rescaling the variables with the magnetic length $\ell_c$ as
\begin{equation}
h_{2,4} \to    h_{2,4}\; \frac{J}{\pi\ell_c^2}~~~~,~~~~r^2\to r^2\;\pi\ell_c^2 \; ,
\label{rescaling}
\end{equation}
we find the adimensional equation,
\begin{equation}
\frac{d^2\theta}{dr^2}+\frac{1}{r}\frac{d\theta}{dr}+2 h_2\sin(2\theta)-4 h_4\sin(4\theta)=0 \; ,
\end{equation}
with the boundary conditions $\lim_{r\to\infty}\theta(r)=0$ and $\lim_{r\to\infty}\theta'(r)=0$.

To solve this equation we transform the boundary value problem into
an initial condition problem and use the shooting method to seek for
solutions. In this method one shoots the initial derivative,
integrates the equation with a Runge-Kutta method of order 4 and,
according to the accuracy to match the boundary values,
the initial conditions are corrected. This procedure is iterated
looking for convergence.

To actually solve the equation, we need to fix $h_2$ and $h_4$.
There is by now extended experimental work in different samples and
regions of magnetic field and density that clearly shows that, while
the anisotropy appears around $T=150 mK$, the energy scale of the
aligning potential is about $1 mK$ per electron. In our model, this
is compatible with values of  $h_2$ and $h_4$ of the order of
$10^{-2}$ in units of $J/\pi\ell_c^2$. For instance, in ref.\
\onlinecite{Manusakis1}, the isotropic/anisotropic transition at
$\nu=9/2$ was successfully fitted with a value of $h_2=0.05$. Of
course, the specific value may changed for different
samples and for different filling factors and in-plane magnetic
fields.  As we have stated before, due to the bidimensionality, the
prefactor in the decay rate, eq.\  (\ref{Gamma}), depends just on
the energy difference between the minima, say $2h_2$, and
not on the height of the barrier, proportional to $h_4$. This means
that for a given value of the  decay time, the energy of the
critical bubble only has a logarithmic dependence on $h_2$.
Then, what actually  matters for the calculation of the energy
of the critical bubble is the order of $h_2$ and not its precise
value. For this reason, we will fix a reasonable  value of $h_2$ and
we will make the numerical calculations for several values of $h_4$
in the range of $10^{-2}$. When comparing with experiments, we will
comment again on these values and will show the robustness of the
results for small changes of these quantities.

\begin{figure}
  \centering
\includegraphics[width=0.45\textwidth]{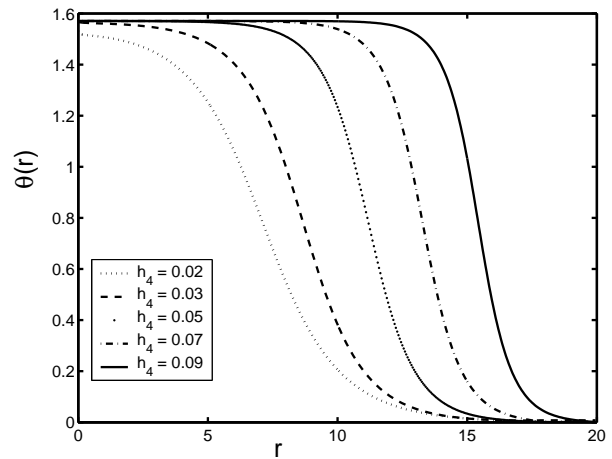}
 \caption{Bubble profiles for different values of $h_4$. We have fixed $h_2=0.02$ for all solutions. $r$ is measured in units of $\sqrt{\pi}\ell_c$ and $h_4$ in units of $J/\pi\ell_c^2$.}
  \label{bubbles}
\end{figure}

\begin{figure}
  \centering
\includegraphics[width=0.46\textwidth]{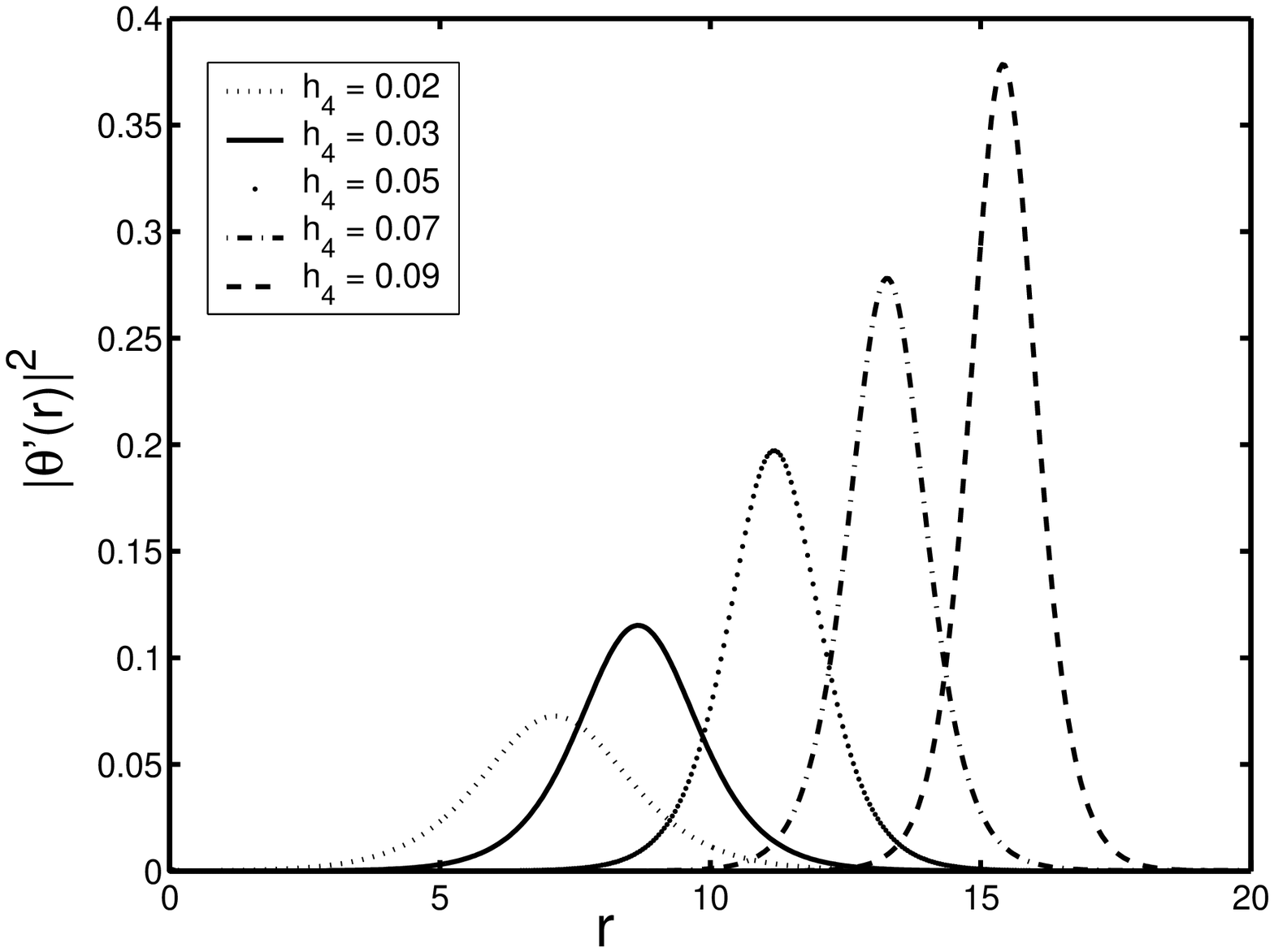}
 \caption{Wall profiles for different values of $h_4$. We have fixed $h_2=0.02$ for all solutions. $r$ is measured in units of $\sqrt{\pi}\ell_c$ and $h_4$ in units of $J/\pi\ell_c^2$.}
  \label{walls}
\end{figure}

\begin{figure}
  \centering
\includegraphics[width=0.47\textwidth]{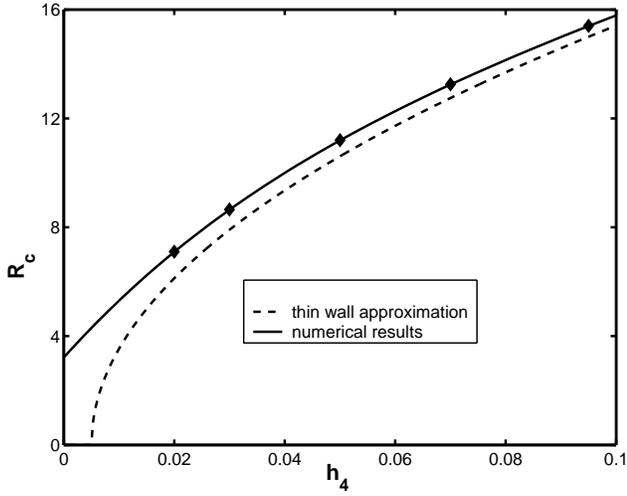}
 \caption{Critical radius as a function of $h_4$. The continuous line is a polynomial fit.  
 The dash line is the critical radius in the thin wall approximation calculated with eq.~(\ref{eqRc}).  
 We have fixed $h_2=0.02$. $R_c$ is measured in units of $\sqrt{\pi}\ell_c$ and $h_4$ in units of $J/\pi\ell_c^2$.}
  \label{Rc}
\end{figure}

In fig.~\ref{bubbles}  we show a set of solutions with fixed
$h_2=0.02$ for several values of $h_4$ while in fig.~\ref{walls} the
square modulus of the derivatives are depicted for the same values
of the parameters. Firstly, we observe the general features of a
critical bubble described in the previous sections as expected. All
solutions begin near $\theta=\pi/2$ and asymptotically go
to $\theta=0$. The critical radius $R_c$ can be read from the
position of the maxima of fig.~\ref{walls} and we can estimate the
wall thickness as the width of peaks at some arbitrary fixed height
(half the peak height, for instance).  According to the
previous section, we see that $R_c$ is an increasing function of
$h_4$ and, the greater $h_4$ the smaller the wall thickness. We
collect this information in fig.~\ref{Rc}, where we draw the
critical radius as a function of $h_4$. The continuous line
connecting the points is a polynomial fit. We also draw with a
dash line  the function $R_c(h_4)$ given by
eq.~(\ref{eqRc}). Interestingly, except for very low values of
$h_4$, the variational result reasonable agrees with the exact
result; the greater $h_4$ the better the approximation.

\begin{figure}
  \centering
\includegraphics[width=0.45\textwidth]{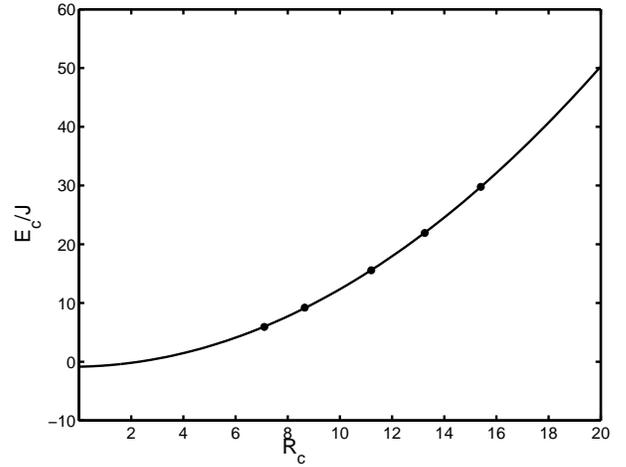}
 \caption{Critical energy in units of $J$ as a function of the critical radius $R_c$ expressed in units of $\sqrt{\pi}\ell_c$. The continuous line is a polynomial fit of order two.}
  \label{RcEner}
\end{figure}

Finally, we compute the energy of each critical bubble by
numerically integrating eq.~(\ref{BubbleEnergy}). In
fig.~\ref{RcEner}, we show the energy as a function of the critical
radius. The continuous line is a polynomial fit of second order.
Note that we are perfectly fitting five points with a second order
polynomial. The reason for that is simple:  the critical energy can
be cast in terms of the critical radius as $E_c=\Delta{\cal F}\; \pi
R_c^2$. We have estimated this parameter to be $\Delta{\cal F}=2
h_2$.  Therefore, we expect a quadratic dependence of the form $E_c=
a_2  R_c^2$, with the quadratic coefficient given by $a_2=2 h_2
\pi$. We have found a fitting value $a_2=0.1257$, in excellent
agreement with the thin wall estimation with the value $h_2=0.02$,
fixed for all the critical bubbles. It seems striking to compare
figures \ref{Rc} and \ref{RcEner}. While in  fig.~\ref{RcEner} the
exact results match thin wall calculations almost perfectly, in
fig.~\ref{Rc} we see a clear deviation. The reason is that in
determining  $\Delta{\cal F}$, the only contribution comes from the
second term of eq.~(\ref{BubbleEnergy}). The contribution of order
$R_c^2$ essentially comes from the constant part of the
bubble profile and our variational function estimates quite well
this area. Indeed, this is the only source of error in the results
in fig.~\ref{RcEner} and, for this reason, the numerical
calculations and the variational estimations match almost perfectly.
On the other hand, for the estimation  of $\sigma$, necessary to
evaluate the critical radius (see eq.~(\ref{eqRc})), we
have two contributions (see eq.~(\ref{sigmaxi})). Both of them
come from the bubble wall and, as we can see from
fig.~\ref{fig:ansatz}, our variational ansatz is not so good in that
region. However, as it is shown in fig.~\ref{Rc}, the approximation
gets better as the value of $h_4$ grows, since the barrier height
increases and, consequently, the wall gets thiner in this limit.

\section{Experiment Interpretation \label{experiments}}

In this section we analyze the experimental results of
ref.~\onlinecite{Eisenstein} in the context of our model of nematic
nucleation. The main purpose of this section is to analyze whether
the above mention data can be interpreted as thermal activation over
a barrier and if the homogeneous nucleation theory of Langer is
applicable. Moreover, we want to check the idea of a  nematic liquid
proposed earlier\cite{F-K,Manusakis1} to describe  the ground state
properties of these systems. To reach this aim, we need to make not
only qualitative comparisons with experiment, but concrete
predictions, specially about the size of the nematic domains and
walls, in order to further check this picture.

In ref. \onlinecite{Eisenstein}, Cooper {\em et.\  al.\ } showed
clear evidences of metastable behavior in the resistance anisotropy
orientation of very clean 2DEG. The measurements were done in a
square sample of area $25 mm^2$ and density $N_s=3\times 10^{11}
cm^{-2}$. The main result was that the ``hard''  and
``easy'' directions of the longitudinal resistance depend, at half
filling, $\nu=9/2,11/2,13/2\ldots$,  on the magnetic field sweep.
A typical hysteresis diagram associated with
metastability was shown and,  in the same work, the authors were be able 
to ``quench'' the state into a metastable direction and to follow the
slow relaxation to the equilibrium. The decay rate strongly
depended on temperature and on filling factor.

The procedure for analyzing  the experiment in the context of our
model is the following: we take three measured quantities
from the experiment, the time needed to complete the anisotropy
``rotation'', the   equilibrium temperature at which the decay was
observed and the critical temperature for the isotropic/anisotropic
transition. With this input, we  compute the energy of the critical
bubble by inverting eq. (\ref{tau}).  Note that this
calculation is model independent and its result gives us information
about the applicability of Langer theory. With this value of the
Energy we estimate the critical radius $R_c$ from fig.\
\ref{RcEner}. The value of  $h_2$, as was explained before,
was taken from several previous experiments and theoretical
fittings, and the value of the new parameter $h_4$ is predicted form fig.\ \ref{Rc}.

In table~\ref{table} we have collected the experimental results for  $\nu=13/2$ in the first two columns. $\tau$ is an estimation of the typical time to complete the transition and $T$ is the equilibrium temperature at which the time evolution was witness.

Assuming that the main mechanism is thermal activation, it is
immediate to estimate, from this data, the value of the critical
energy by using eq.~(\ref{tau}). The area  was taken from the
experiment and we have fixed the value of $h_2=0.02 J/(\pi
\ell_c^2)$. Due to the logarithmic dependence, the critical energy
is not sensible to this particular value but just to its order. We
have fixed it by considering that the strength of the
native symmetry breaking potential is about $1 mK$ per
electron\cite{{native}}. The results are depicted in the third
column of table~\ref{table}.

The first observation is that $E_c/T$ takes values  between $40$ and $50$. Therefore, the homogeneous nucleation theory of Langer is a reasonable approximation, at least for first order estimations since $E_c/T>>1$. Note that this value of the energy corresponds to $E_c\sim 3 K$.

The next step is to compute $E_c/J$. The stiffness $J$ can be
considered of the same order of the isotropic/anisotropic
transition, $J= \alpha\; T_c$, where $\alpha$ is a constant of order
one and $T_c$ is the critical temperature for the
isotropic/anisotropic transition. While  $T_c$ is an experimental
data, $\alpha$ should be computed from the
model. The order of $T_c\sim 150 mK$ is not difficult to obtain;  a
more accurate number  is generally more involved since the
transitions are rounded by disorder and other effects. On the other
hand, the value of $\alpha$ for our model is  not simple to compute.
When ignoring the small symmetry breaking potential, the
renormalization group analysis of the Kosterlitz-Thouless
transition\cite{Wexler} gives an estimation of $\alpha\sim 3$.
Moreover, Monte-Carlo simulations  of the full
model\cite{Manusakis1}, fitting the experimental data of
ref.~\onlinecite{QHSExp} (at $\nu=9/2$), give a reduced value of
$\alpha\sim 1.1$. For the estimation of $E_c/J$ we have used
$T_c\sim 150 mK$, which is reasonable for $\nu=13/2$ (note
in ref.~\onlinecite{Eisenstein} that at $100mK$ the anisotropy is
completely developed at this filling factor), and the value of
$\alpha$ obtained from a renormalization group analysis. These
values are shown in the fourth column of table~\ref{table}.

Now we are ready to predict the value of the critical radius and the
parameter $h_4$ by using the numerical calculations of
figures \ref{RcEner} and \ref{Rc} respectively. We show these
results in the last two columns of table \ref{table}.

We find for the critical
radius $R_c\sim 2.6 \mu m $, where we have considered  the magnetic length $\ell_c= 197
\stackrel{o}{\rm A}$ (at $\nu=13/2$). By taking into account the
electronic density of the sample, this value gives an estimation of
$10^4$ electrons inside the critical bubble. We see that this
prediction is completely reasonable since, the dimension of the
critical domains is  big enough to approximately contain ten
broken stripes in each domain, provided we consider the Hartree-Fock
value of the stripe period\cite{Fogler} as a reasonable estimation.
Moreover, the domains are $10^6$ times smaller than the size of the
sample.

Although the precise value of $h_2$ was fixed by hand, the size of the
critical domain predicted is not very sensible to that value. In
fact, since the critical energy is only logarithmically dependent on
$h_2$ and $E_c=\pi\Delta{\cal F} R_c^2$, we see from eq.
\ref{deltaF} that $R_c\sim 1/\sqrt{h_2}$. That means that
when varying $h_2$ over a range of reasonable values, the
critical radius only changes up to $15\%$. We also note
that the predicted value of $h_4$ is almost equal to $h_2$ and shows a very tiny
temperature dependence. Recently, it was pointed out\cite{piezo}
that piezoelectricity in $GaAl$ can induce an aligning potential of
the form $\cos(4\theta)$. In that work, it was predicted that the
barrier between the two degenerated minima is $10^{-4}$ times the
Coulomb energy, roughly $1 mK$ per electron. This value is in
complete agreement with the values of $h_4$ predicted in this paper.

We have repeated the calculations with different values of $h_2$ and
we have found a linear relation  $h_2/h_4\sim O(1)$. This is
compatible with eq. (\ref{Ecthinwall}) which, for a fixed
value of $E_c$, predicts the linear dependence in the thin
wall approximation (up to logarithmic corrections). Of course, in
order to determine a precise value for the parameter $h_2$ it is
necessary to fit the data for all values of temperatures above the
critical temperature.

It is important to note that, although the homogeneous nucleation approximation seems to be a reasonable one, the thin wall approximation is not accurate for the regime of this experiment. This fact can be observed in fig.~\ref{Rc}. By using eq.~(\ref{thinwall}) we estimate $\xi/R_c\sim 2/3$. Then, the domain walls are not sharply defined getting broader on a smooth interphase of approximately $1\mu m$.

\begin{table}
 \begin{tabular}{||c|c|c|c|c|c|c||}
 \hline
{\bf  $T$ [mK] }& {\bf $\tau$ [s]} & {\bf $E_c/T $} &  {\bf $E_c/J$ }& {\bf $R_c/\sqrt{\pi}\ell_c$} &{\bf  $h_4$} \\
\hline
   50       &   $3.6\times 10^{4}$       &  51      &     5.7        &   6.8       &    0.020    \\
 \hline
 70       &   $ 6\times 10^{2}$          &  47      &    7.3        &   7.6       &    0.023    \\
 \hline
 90       &   1                          &  41       &   8.2        &   8.1       &    0.026    \\
 \hline
 \end{tabular}
\caption{Summary of experimental results and theoretical predictions. $T$ is the equilibrium temperature of the final state, $\tau$ is the typical time that takes the transition to be completed. $E_c$ is the energy of the critical bubble, $R_c$ is the critical radius, $\ell_c$ the magnetic length and
$h_4$ the tetragonal component of the native potential measured in units of $J/\pi \ell_c^2$.}
\label{table}
\end{table}

\section{Summary and discussions\label{summary}}

We have analyzed the experimental
observation\cite{Eisenstein}  that the native
potential, responsible for the resistance anisotropy alignment in
quantum Hall fluids at half filling, has a non-trivial structure,
which favors two orthogonal directions.

In this framework, we have studied an $XY$ model, which describes 
a nematic fluid in an external symmetry breaking
potential, compatible with the nematic symmetry. We have considered
the tetragonal coefficient  of the Fourier expansion of the
potential  and we have shown how it can  produce two orthogonal
local minima structure.

In this picture, we have assumed that the ``rotation'', observed in
the anisotropy axis, is mainly driven by thermal activation over a
barrier. Then, the decay of a metastable direction
into the orthogonal direction, which represents the true ground
state, is dominated by nucleation of nematic domains.  

To compute decay rates we have used the homogeneous nucleation theory\cite{Langer} and, after comparing with experiment, we have concluded that it is a reasonable approximation for first order estimations since $E_c/T\sim 50$ for the data of ref.~\onlinecite{Eisenstein}.

We have implemented an analytical variational approach inspired in
the bubble solutions of a quartic potential and we have computed the
critical energy and radius in the thin wall
approximation. In order to check the quality and the range of
applicability of our approximations, we have numerically integrated
the critical bubble differential equation, and we have computed its
energy for a wide range of parameters.

By comparing with the experiment, we were able to predict the radius
of the nematic critical domains of the order of $R_c\sim  2.6 \mu
m$; each domain  approximately contains $10^4$ electrons. However, we
found that the width of the wall is quite broad. There is a smooth
transition from the true ground state to the metastable state spread
in a region of the order of $\xi\sim 1\mu m$. Therefore, although
the mechanism of homogeneous nucleation  seems to be a reasonable
assumption, the thin wall approximation is not accurate for the
actual regime showed in the data.

The prediction on the size of the critical domains is robust in the
sense that it does not strongly depend on the detailed value
of $h_2$. On the other hand, it is quite sensible to the
experimental determination of the critical temperature $T_c$ for the
isotropic/anisotropic phase transition and to the computation of the
stiffness of the $XY$ model in the presence of an external field. A
Monte-Carlo fitting of the complete data, including high
temperatures around $T_c$ for different filling factors,  would help
to improve the accuracy of the predictions of $R_c$ and $h_4$.

Finally, we would like to point out that, while the
mechanism proposed in this paper is in agreement with experiment, it
is not the only possibility for its explanation. Indeed, the origin
of metastability could be produced from competition between
different crystal liquid phases rather than from a native potential
or, perhaps, from combinations of both effects. In a recent
paper\cite{nematic-hexatic}, it was shown that the competition
between nematic and hexatic order parameter leads to a first order
phase transition with a clear signature of metastability in the
orientation of the principal axis. The quantum hexatic phase could
be produced from melting of the crystal bubble phase. All these
anisotropic states are in fact competing close to half filling,
contributing to  the interesting complex structures of longitudinal
resistivity reported.  We hope to report on this possibility soon.

\begin{acknowledgments}
The Brazilian agencies {\em Conselho Nacional de Desenvolvimento Cient\'{\i}fico e Tecnol\'{o}gico (CNPq)},
and the {\em Funda{\c {c}}{\~{a}}o de Amparo {\`{a}} Pesquisa do Estado do Rio de Janeiro (FAPERJ)}  are acknowledged for the financial support.
This work was partially supported by the bi-national collaboration UERJ-Brazil/ICIMAF-Cuba, founded by {\em Centro Latino-Americano de F\'\i sica, CLAF}.
\end{acknowledgments}

\end{document}